\documentclass[preprint, superscriptaddress,amsmath, nofootinbib]{revtex4-1}
\usepackage{graphicx}
\usepackage{latexsym}
\usepackage{slashed}
\usepackage{bm}
\usepackage{color}
\usepackage[colorlinks,citecolor=blue,urlcolor=blue,linkcolor=red]{hyperref}
\usepackage{diagbox}
\usepackage{makecell}
\pdfoutput=1
\usepackage{dcolumn}
\usepackage{bm}
\usepackage{multirow}

\begin{document}

\title{Probing the Gauge-boson Couplings of Axion-like Particle at 
the LHC and High-Luminosity LHC}
\def\slash#1{#1\!\!\!/}

\author{Kingman Cheung,}
\email{cheung@phys.nthu.edu.tw}
\affiliation{Department of Physics, National Tsing Hua University, Hsinchu 30013, Taiwan} 
\affiliation{Center for Theory and Computation,
National Tsing Hua University, Hsinchu 30013, Taiwan}
\affiliation{Division of Quantum Phases and Devices, School of Physics, Konkuk University, Seoul 143-701, Republic of Korea}

\author{Wanyon Hsiao}
\email{mark0706@gmail.com}
\affiliation{Department of Physics, National Tsing Hua University, Hsinchu 30013, Taiwan} 
\affiliation{Center for Theory and Computation,
National Tsing Hua University, Hsinchu 30013, Taiwan}

\author{C.J. Ouseph}
\email{ouseph444@gmail.com}
\affiliation{Department of Physics, National Tsing Hua University, Hsinchu 30013, Taiwan} 
\affiliation{Center for Theory and Computation,
National Tsing Hua University, Hsinchu 30013, Taiwan}

\author{Chen Wang}
\email{elian890415383@gmail.com}
\affiliation{Department of Physics, National Tsing Hua University, Hsinchu 30013, Taiwan} 
\affiliation{Center for Theory and Computation,
National Tsing Hua University, Hsinchu 30013, Taiwan}

\date{\today}

\begin{abstract}

In this work, we calculate the sensitivities on the gauge-boson couplings
$g_{aZZ}$, $g_{aZ\gamma}$, and $g_{aWW}$ of an axion-like particle (ALP)
that one can achieve at the LHC with $\sqrt{s}=14$ TeV and 
integrated luminosities of 300 fb$^{-1}$ (current run)
and 3000 fb$^{-1}$ (High-Luminosity LHC).
We focus on the associated production processes $pp\to Za \to (l^+l^-)(\gamma\gamma)$ 
and $pp\to W^\pm a \to (l^\pm \nu)(\gamma\gamma)$.
We show that better sensitivities on these gauge couplings can be achieved at the LHC
for $M_a = 1-100$ GeV, down to the level of $10^{-4}\,{\rm GeV}^{-1}$. In conclusion, this study emphasizes the significance of the investigated channels in constraining 
the ALP couplings at the LHC, offering valuable insights for future 
experiments dedicated to ALP detection.

\end{abstract}

\maketitle

\section{Introduction}\label{sec:intro}

One of the long-standing problems in the standard model (SM) is the strong CP problem 
\cite{Peccei:1977hh}. It arises from the term $\theta G_{\mu\nu}\tilde{G}^{\mu\nu}$ in QCD, 
in which $\theta \sim O(1)$. Such a term 
contributes substantially to the neutron electric-dipole moment (EDM) 
$d_n (\theta) \approx 2.4\times 10^{-16}\,\theta\, e\cdot{\rm cm}$.
Nevertheless, non-observation of the neutron EDM gives an upper limit 
$|d_n| <  1.8 \times 10^{-26}\,e\cdot{\rm cm}$ \cite{Abel:2020pzs}, which
requires $\theta < 10^{-10}$.
Such a small coefficient in the Lagrangian is unnatural, which is
coined as the strong CP problem. 
One of the best solutions is by introducing a global Peccei-Quinn (PQ) symmetry $U(1)_{PQ}$ symmetry
\cite{Peccei:1977hh}, which was spontaneously broken 
by a dynamical axion field. The resulting pseudo-Nambu-Goldstone boson is known as the QCD axion \cite{Peccei:1977hh,Weinberg:1977ma,Wilczek:1977pj}. 
The neutron EDM constraint demands the breaking scale of the PQ symmetry to be 
very high with $f_a > 10^9$ GeV, implying a tiny mass to the axion and very small 
couplings to the SM particles. 

The axion can also serve as a dark matter candidate \cite{Preskill:1982cy,Abbott:1982af,Dine:1982ah}.
If we do not require the pseudo-Nambu-Goldstone boson to be the solution of the strong CP problem, 
the mass of the axion is not restricted by the breaking scale $f_a$. 
Such a hypothetical particle, called axion-like particle (ALP), is also a pseudoscalar boson.
The ALP has a much wider range of mass and couplings that it can serve as the dark matter candidate.
The axion mass and couplings to SM particles can extend over many orders of magnitude, which are
constrained by astrophysical and cosmological observations, as well as collider experiments
(for a comprehensive summary of constraints please see {\tt https://cajohare.github.io/AxionLimits/}).

The ALP as a dark matter candidate is not the motivation of this work,
unless the couplings of the ALP are extremely small such that the lifetime 
is longer than the age of the Universe.
On the other hand, the ALP as a low-scale inflaton is another possibility
that the ALP can decay \cite{Takahashi:2018tdu}. 
In some 
models~\cite{CidVidal:2018blh,Acanfora:2023gzr,Fitzpatrick:2023xks,Dror:2023fyd,Ghosh:2023tyz,Armando:2023zwz}, 
the axion or ALP can serve as the mediator to the dark-matter sector 
regardless of its mass or lifetime.
In this work, we consider the potential sensitivities on the parameter space of the ALP model
that one can achieve at the current LHC (${\cal L}= 300\;{\rm fb}^{-1}$)
and the future High-Luminosity LHC (${\cal L}= 3000\;{\rm fb}^{-1}$). 
In this work, we focus on the gauge couplings
$g_{aZZ}$, $g_{aWW}$, and $g_{aZ\gamma}$ of the axion $a$. 
In principle, due to gauge invariance, it can also lead to the sensitivity 
on $g_{a\gamma\gamma}$.  Instead, we obtain the sensitivities in 
a model-independent manner.

In this work, we focus on the associated production of the axion $a$ with
a $Z$ or $W$ boson, followed by the leptonic decay of the $Z$ or $W$ boson and 
the decay of $a\to \gamma\gamma$, i.e.,
\[
 pp \to Z a \to (l^+ l^-) (\gamma \gamma)
\]
and
\[
 pp \to W^\pm a \to (l^\pm \nu) (\gamma \gamma) \;,
\]
where $l = e, \mu$. 
The associated production $pp\to Za$ can proceed via a $Z$ or $\gamma$ propagator,
which can then probe $g_{aZZ}$ and $g_{aZ\gamma}$, respectively. 
On the other hand, the process $pp \to W^\pm a $ probes $g_{aWW}$.
We obtain the sensitivities on these gauge couplings for $M_a$ from 1 to 100 GeV.
In the analysis, we found that when $M_a \leq 25$ GeV, the two photons from 
axion decay are quite close to each other, which form, what we called, a photon-jet.
While for $M_a \geq 25$ GeV the two photons from axion decay are well separated.
Therefore, we choose different selection procedures for low-mass $M_a$ and high-mass
$M_a$. More details are given in Sec.~\ref{sec:Exp}.

The organization is as follows. In the next section, we describe the 
relevant interactions of the ALP. In Sec.~\ref{sec:Constrints}, we 
summarize the existing constraints on $g_{aZZ}$, $g_{aZ\gamma}$, and $g_{aWW}$. 
In Sec.~\ref{sec:Exp}, we discuss in details the signal and background 
analysis. In Sec.~\ref{sec:results}, we show our results. We conclude in
Sec.~\ref{sec:conclusion}. 

There have been a number of studies to explore the properties of 
ALPs at different colliders~\cite{Mimasu:2014nea,Jaeckel:2015jla, Bauer:2017ris,Bauer:2020jbp,Chala:2020wvs,Bonilla:2021ufe,Aiko:2024xiv}.

\section{The Model}\label{sec:Model}
The axion, as a pseudo-Goldstone boson, has derivative couplings to fermions, as well
as $CP$-odd couplings to the gauge field strengths. Before rotating 
the $B$ and $W^i$ fields to
the physical $\gamma,~Z,~W^\pm$, the interactions of the 
axion are given by following equations: \cite{Brivio:2017ije,Georgi:1986df,Ren:2021prq}
\begin{equation}\label{Eq.1}
\mathcal{L}=\mathcal{L}_f+\mathcal{L}_{g}+\mathcal{L}_{BB}+\mathcal{L}_{WW}
\end{equation}
where
\begin{equation}
    \mathcal{L}_f=-\frac{ia}{f_a}\sum_{f}g_{af}~m^{diag}_f\bar{f}\gamma_5f
\end{equation}
\begin{equation}
    \mathcal{L}_g = -C_g\frac{a}{f_a}G^A_{\mu\nu}\Tilde{G}^{\mu\nu,A}
\end{equation}
\begin{equation}
    \mathcal{L}_{BB} = -C_{BB}\frac{a}{f_a}B_{\mu\nu}\Tilde{B}^{\mu\nu}
\end{equation}
\begin{equation}
    \mathcal{L}_{WW} = -C_{WW}\frac{a}{f_a}W^i_{\mu\nu}\Tilde{W}^{\mu\nu,i}.
\end{equation}
where $a$ represents the ALP field, $f_a$ is the ALP decay constant, $A=1,....8$ is the $SU(3)$ color index and $i=1,2,3$ is the $SU(2)$ index. The term ${\cal L}_f$ describes the fermionic couplings of the ALP.
In principle, there can be non-trivial flavor structures in the ALP couplings such as $g_{ij} \frac{(\partial_\mu a) } {f_a} (\bar f_i \gamma^\mu \gamma^5 f_j)$, where $i,j$ denote the generation indices. Nevertheless, we set them all to zero in our study. The $B,W^3$ fields are rotated into $\gamma, Z$ by
\begin{equation}\label{Eq.2}
\begin{pmatrix}
W^3_\mu \\
B_\mu 
\end{pmatrix}= \begin{pmatrix}
c_w & s_w \\
-s_w & c_w
\end{pmatrix} \begin{pmatrix}
Z_\mu \\
A_\mu 
\end{pmatrix},
\end{equation}
where $c_w$,$s_w$ are cosine and sine of the Weinberg angle. 
The axion interactions with the fermions and the physical gauge bosons are given by
\begin{equation}\label{Eq.3}
\begin{split}
\mathcal{L}=-\frac{ia}{f_a}\sum_{f}g_{af}m^{diag}_f\bar{f}\gamma_5f-C_g\frac{a}{f_a}G_{\mu\nu}^A\tilde{G}^{\mu\nu A}-\frac{a}{f_a}\big[(C_{BB}c^2_w+C_{WW}s^2_w)F_{\mu\nu}\tilde{F}_{\mu\nu}+&\\(C_{BB}s^2_w+C_{WW}c^2_w)Z_{\mu\nu}\tilde{Z}_{\mu\nu}+2(C_{WW}-C_{BB})c_ws_wF_{\mu\nu}\tilde{Z}_{\mu\nu}+C_{WW} W^+_{\mu\nu} \tilde{W}^{- \mu\nu}\big] \;.
\end{split}
\end{equation}
The dimensionful couplings associated with ALP interactions from Eq. (\ref{Eq.3}) are given by:
\begin{equation}\label{Eq.4}
g_{a\gamma\gamma}=\frac{4}{f_a}(C_{BB}c_w^2+C_{WW}s_w^2),
\end{equation}
\begin{equation}\label{Eq.5}
g_{aWW}=\frac{4}{f_a}C_{WW},
\end{equation}
\begin{equation}\label{Eq.6}
g_{aZZ}=\frac{4}{f_a}(C_{BB}s_w^2+C_{WW}c_w^2),
\end{equation}
\begin{equation}\label{Eq.7}
g_{aZ\gamma }=\frac{8}{f_a}s_wc_w(C_{WW}-C_{BB}) \,.
\end{equation}
Note that $g_{a\gamma\gamma}$, $g_{aZZ}$, $g_{aWW}$, and $g_{aZ\gamma}$ are not independent if we 
assume the $SU(2)$ symmetry relation as in Eq.~(\ref{Eq.1}). By choosing $O(1)$ coefficients for
$C_{WW}$ and $C_{BB}$ we can convert the existing constraints on $g_{a\gamma\gamma}$ to 
the others.

\section{Existing Constraints on ALPs }\label{sec:Constrints}
   In this section, we discuss the existing constraints on the ALP-weak gauge boson couplings.
   Given that we focus on the mass range $1~\text{GeV} \leq M_a \leq 100~\text{GeV}$, 
   we summarize the constraints for this ALP mass range. 
   Various ALP-photon coupling limits have been established in various collider 
   experiments \cite{L3:1992kcg, CidVidal:2018blh, Mariotti:2017vtv, Mosala:2023sse,Aiko:2023trb,
Aiko:2024xiv, Liu:2023bby}. These limits 
   can be converted into $g_{aZZ}$, $g_{aZ\gamma}$, and $g_{aWW}$ using Eqs.~(\ref{Eq.4})--
   (\ref{Eq.7}) by choosing $O(1)$ coefficients for
$C_{WW}$ and $C_{BB}$. The corresponding plots are presented in Fig.~\ref{fig:final_result}, 
   labeled as {\tt "photons (various)"}.
The label {\tt "photons (various)"} includes the following: the bound established by the L3 collaboration, focusing on hadronic final states accompanied by a hard photon, was surpassed by exclusions from the LHC~\cite{L3:1992kcg}. The exclusion for the ``Flavour" region was based on data from Babar and LHCb~\cite{BaBar:2011kau,Benson:2018vya}. At high axion masses near the TeV scale, the LHC bounds exceeded those from LEP due to enhanced 
axion production via gluon-gluon fusion. Strong limits are obtained from run 1 
data~\cite{Jaeckel:2012yz,Mariotti:2017vtv}.

There are other limits on the $g_{a\gamma\gamma}$ in additional to those included in "photon (various)"
in Fig.~\ref{fig:final_result}. Both CMS~\cite{CMS:2018erd} and ATLAS~\cite{ATLAS:2020hii},
based on the $\gamma\gamma \rightarrow a \rightarrow \gamma\gamma$ process in PbPb collisions, 
constrained $g_{a\gamma\gamma}$ or $1/f_a \leq  0.1 -1 \,{\rm TeV}^{-1}$ for $M_a \sim 10 - 100$ GeV.
On the other hand, the CMS and TOTEM ~\cite{CMS:2023jgd}, using diphoton production, excluded 
$1/f_a > 0.03 - 1 \, \text{TeV}^{-1}$ for $M_a = 500 - 2000$ GeV.
Another search  by ATLAS based on light-by-light scattering mediated by an ALP constrained 
$1/f_a < 0.04-0.09 \, {\rm TeV}^{-1}$ for $M_a = 150 - 1600$ GeV~\cite{ATLAS:2023zfc}.
If we were to convert these constraints on $g_{a\gamma\gamma}$ to $g_{aZZ, aZ\gamma}$ 
based on gauge invariance, it would be similar to the shaded regions in Fig.~\ref{fig:final_result}.
On the other hand, 
there are also constraints on the flavor-changing neutral current (FCNC) interactions of the ALP
from the kaon and B-meson rare decays in invisible ALP and visible ALP channels 
\cite{Gavela:2019wzg,Bauer:2021mvw}. The invisible ALP channel can contribute to
$K \to \pi \nu \bar{\nu}$ while the visible ALP channel can give rise to a 
displaced vertex in the decays
such as $B \rightarrow K^* \mu^+ \mu^-$. The constraint is stronger with 
$C_{WW}/f_a < 0.2-0.3 \,{\rm TeV}^{-1}$ for lighter $M_a \leq 0.1$ GeV from kaon decays.  
For $M_a =0.1 - 5$ GeV the constraint is roughly $C_{WW}/f_a \leq 1-10\, {\rm TeV}^{-1}$ from 
$B \rightarrow K^* \mu^+ \mu^-$. These constraints are weaker than the shaded regions shown in
Fig.~\ref{fig:final_result}.

\subsection{Limits on $g_{aZZ}$ versus $M_a$}\label{sec:Constrints_01}
    In Ref.~\cite{Craig:2018kne}, the process $pp \to \rm{triboson}$ at the LHC was investigated 
    to explore the limit on $f_a$. By employing Eqs.~(\ref{Eq.5}) -- (\ref{Eq.7}), 
    the constraint on $g_{aZZ}$ can be derived by selecting specific coupling coefficients 
    $C_{WW}$ and $C_{BB}$, labeled as "triboson (LHC)" in Fig.~\ref{fig:final_result}.  
    The CMS Collaboration \cite{CMS:2021} used the ALP-mediated 
    non-resonant $ZZ$ pair production at the LHC and obtained a constraint on $g_{aZZ}$ of approximately $6.6 \times 10^{-4}~\rm{GeV^{-1}}$.  This set of constraints on $g_{aZZ}$ are
    shown in upper-left panel of Fig.~\ref{fig:final_result}.

\subsection{Limits on $g_{aZ\gamma}$ versus $M_a$}\label{sec:Constrints_02}
    The constraint on $g_{aZ\gamma}$ was initially established in Sec. 6.1 of 
    Ref.~\cite{Brivio:2017ije} by considering the uncertainty of the total $Z$ boson width,
    $\Gamma(Z \to \rm{BSM}) \lesssim 2 \rm{MeV}$ at 95\% confidence level. 
    The derived limit was $|g_{aZ\gamma}| < 1.8 ~\rm{TeV^{-1}}$ for ALP masses 
    below the $Z$ boson mass. 
    Another limit was presented in Ref.~\cite{L3:1992kcg} using the 
    $e^+ e^- \to Z \to \gamma + \text{hadron}$ channel, with improvements observed in the region
    $10~\rm{GeV} \lesssim \mathit{M_a} < \mathit{M_Z}$. 
    The constraint on $g_{aZ\gamma}$ can also be implicated from the triboson 
    limit in Ref.~\cite{Craig:2018kne} by selecting $C_{WW} \neq C_{BB}$. 
    A recent work on off-shell ALP production constrained the ALP-weak boson coupling, as 
    demonstrated in Ref.~\cite{Nonresonant_ggF:2021}, further improved the limit on $g_{aZ\gamma}$,
    pushing it down to approximately $4 \times 10^{-4}~\rm{GeV^{-1}}$. They are all summarized 
    in the upper-right panel of Fig.~\ref{fig:final_result}. 

\subsection{Limits on $g_{aWW}$ versus $M_a$}\label{sec:Constrints_03}
    Reference~\cite{Craig:2018kne} also provided the limit on $g_{aWW}$ with the triboson channel,
    estimating the constraint on $g_{aWW}$ to be approximately $10^{-2}~\rm{GeV^{-1}}$ 
    in the large mass region. A more refined result on $g_{aWW}$ was presented in 
    Ref.~\cite{Nonresonant_ggF:2021} to about $5 \times 10^{-4}~\rm{GeV^{-1}}$ in the mass range
    shown.  They are all summarized in the lower panel of Fig.~\ref{fig:final_result}. 

\section{Experiment and Simulation}\label{sec:Exp}

    To simulate the signal events, we utilize the UFO file\footnote{The model file is available for download at \url{https://feynrules.irmp.ucl.ac.be/wiki/ALPsEFT}}~\cite{Brivio:2017ije}, as detailed in the effective Lagrangian presented in Eq.~(\ref{Eq.3}). The generation of parton-level signal and background events is carried out using \texttt{MadGraph5aMC@NLO} \cite{MadGraph:2011} at the leading order. To ensure the accuracy of the simulations, specific cuts are applied during the parton-level event generation, as outlined in the \texttt{run\_card.dat} cuts (we used the default cuts outlined in the \texttt{run\_card.dat}). A total of $10^4$ signal events are generated at the center-of-mass energy $\sqrt{s}=14~\rm TeV$. As for the background events, we generate $10^5$ or $10^6$ for different background channels to maintain the accuracy. More details of background events are discussed in Sec. \ref{sec:Exp_Z_bg} and Sec. \ref{sec:Exp_W_bg}.
    The subsequent steps involve parton showering using \texttt{Pythia8} \cite{Pythia:2007} and 
    detection simulations conducted with \texttt{Delphes3} \cite{Delphes:2013}, 
    incorporating the \texttt{delphes\_card\_ATLAS.tcl} for accuracy and consistency. 
 The basic parton-level selection cuts for photons, charged leptons, and jets are
 \begin{eqnarray}
    &&  p_{T_\gamma} > 10 \,{\rm GeV},\;\;\; |\eta_\gamma| < 2.5\;, \nonumber \\
    &&  p_{T_l} > 10 \,{\rm GeV},\;\;\; |\eta_l| < 2.5\;, \nonumber \\
    &&  p_{T_j} > 20 \,{\rm GeV},\;\;\; |\eta_j| < 2.5\;, R_{\rm cone} = 0.4 \;, \nonumber \\
    &&  \Delta R_{ab} = 0.4 \;\;\;\; {\rm where}\;\;  a,b = \gamma, l, j \;.\nonumber
  \end{eqnarray}


\subsection{$p p \to Z a$ with $Z \to l^+ l^-$ and $a \to \gamma \gamma$}\label{sec:Exp_Z}

    We focus on the ALP that exclusively couples to electroweak gauge bosons with the mass range
    spanning from 1 GeV to 100 GeV. One of the prominent production channels for ALPs at the 
    LHC is the ALP-strahlung process, specifically, 
    $pp\to Za ~(Z\to l^+~l^-), (a\to\gamma\gamma)$, where $l=e, \mu$, as illustrated 
    in the left of Fig.~\ref{fig:ppzax_diagram}.
    \footnote{
        There is a small contribution from $pp \to \gamma^* \to l^+ l^- (\gamma\gamma)$, but it is negligible under the $Z$ peak.}
    Note that for $M_a > M_Z$ the ALP can decay into $Z\gamma$. However, 
    for the choice of $C_{WW}$ and $C_{BB}$ ($g_{aZ\gamma}$ and $g_{a\gamma\gamma}$
    are of similar size) the branching ratio into $a \to Z\gamma$ is of order $10^{-3}$ such
    that the branching ratio of $a \to \gamma\gamma$ is practically 1.
    Also, for the parameter space of $g_{a\gamma\gamma}$ and $M_a$ considered in
    this work, the decay of the ALP is prompt.

\subsubsection{Signal Events}\label{sec:Exp_Z_sig}
  The interaction of the ALP with gauge bosons are listed in Eq.~(\ref{Eq.3}). 
  The parameters chosen for this simulation are set to specific benchmark values:
  $f_a=1~\text{TeV}$, $C_{WW}=2$, $C_{BB}=1$, and $C_g=g_{af}=0$. It is noteworthy that 
  the deliberate assignment of different values to $C_{WW}$ and $C_{BB}$ 
  allows for the presence of the coupling constant $g_{aZ \gamma}$.
  Because of very similar topology in the $q\bar q \to Z^*, \gamma^*
    \to Z a \to (l^+ l^-) (\gamma\gamma)$, the event kinematic distributions for
    other values of $C_{WW}$ and $C_{BB}$ would be very much the same, thus the
    signal selection efficiency is independent of the choice of $C_{WW}$ and
    $C_{BB}$. 

  \begin{figure}
        \centering
        \includegraphics[width=\textwidth]{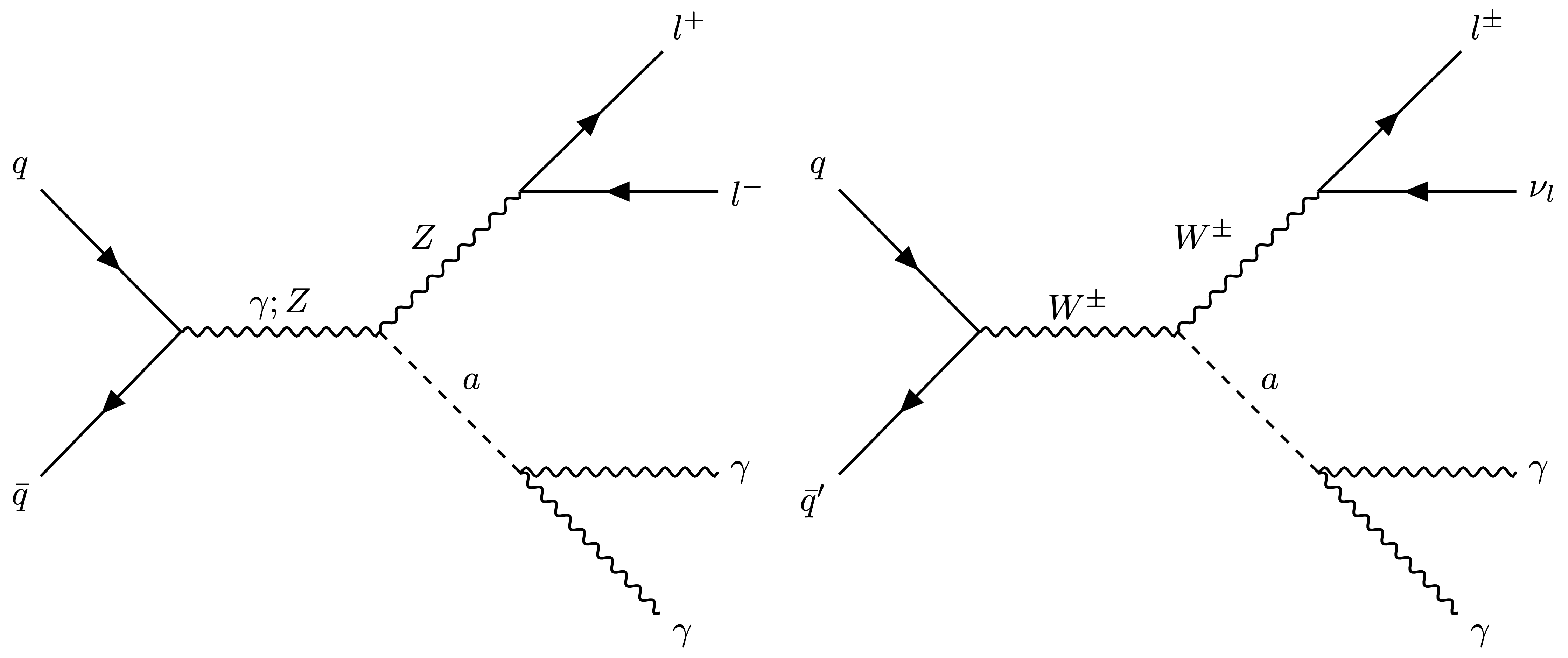}
        \caption{Contributing Feynman diagrams of the signal processes:
        $pp \to Za \to (l^+ l^-) ( \gamma\gamma)$ (left panel) and
     $pp \to W^\pm a \to (l^\pm \nu_l) ( \gamma\gamma)$ (right panel).}
        \label{fig:ppzax_diagram}
    \end{figure}

\begin{figure}
        \centering
        \includegraphics[width=0.75\textwidth]{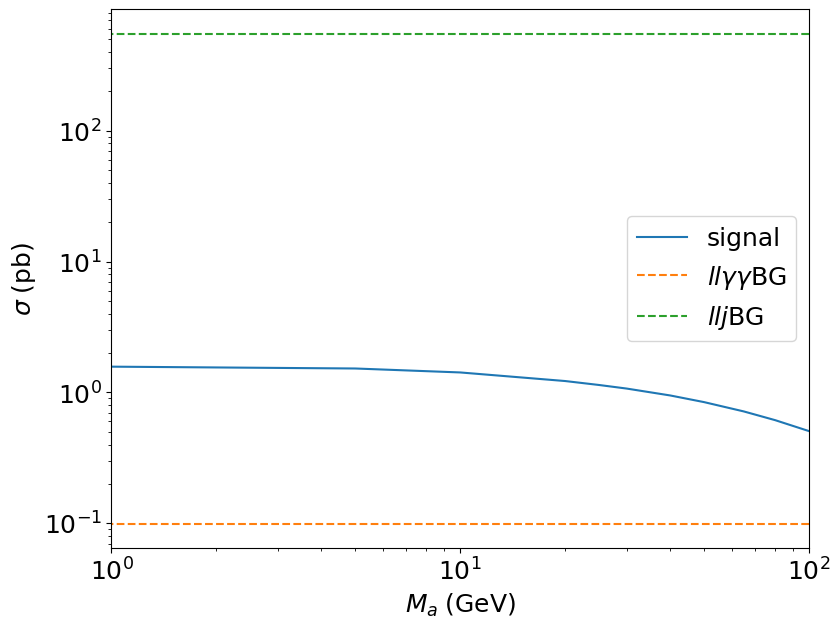}
        \caption{Production cross section for $pp\to Za ~(Z\to l^+~l^-)(a\to\gamma\gamma)$ with
        $ l = e,\mu$ versus $M_a$, including
        the branching ratios at $\sqrt{s} = 14$ TeV. Here we also show the backgrounds
        $ll\gamma\gamma$BG and $llj$BG. }
        \label{fig:ppzax_Xsection}
    \end{figure}

\begin{figure}
        \centering
        \includegraphics[width=\textwidth]{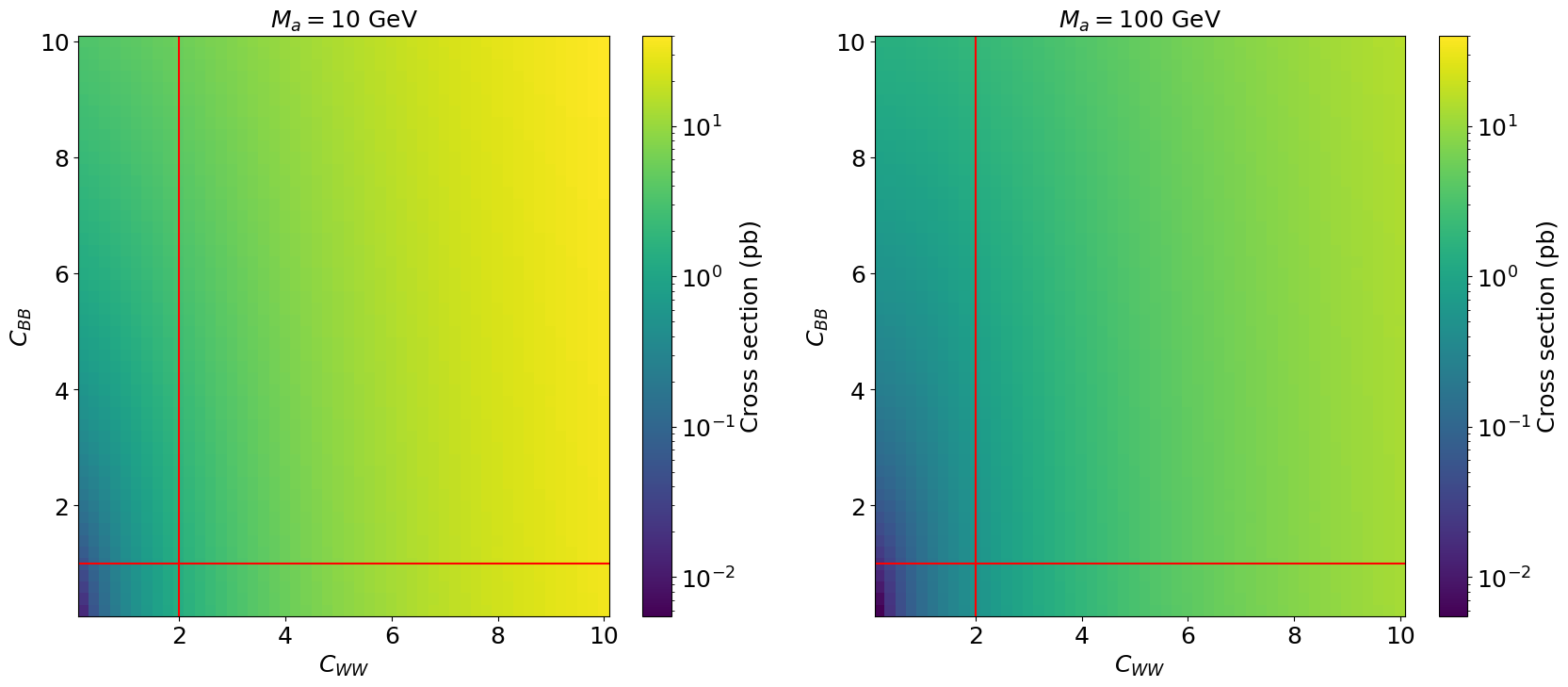}
        \caption{Cross sections for $pp\to Za ~(Z\to l^+~l^-)(a\to\gamma\gamma)$ by 
        varying the parameters $C_{WW}$ and $C_{BB}$ with $\sqrt{s}=14~\text{TeV}$, 
        $f_a = 1~\text{TeV}$ and $M_a=10$ or $100~\text{GeV}$. 
        The cross point of the two red lines represents the benchmark in our study, 
        $C_{WW}=2$ and $C_{BB}=1$.}
        \label{fig:scanCWCB}
    \end{figure}

The mass range for the ALP spans from $1~\rm GeV$ to $100~\rm GeV$. The decay width of the 
ALP is set to ``auto" in \texttt{MadGraph5aMC@NLO}, indicating that the decay width of the ALP
is calculated based on the Lagrangian in Eq.~(\ref{Eq.3}).
Specifically, the leptonic decay modes of electrons and muons are employed for the 
$Z$ boson. Note that although the hadronic and neutrino decay modes could give 
a higher signal event rates, the detection is far more challenging and suffered 
from more background. On the other hand, for the ALP, under our specified benchmark parameter space, the dominant decay mode is a pair of photons.
The corresponding Feynman diagram illustrating the ALP-strahlung process with the decay of the 
$Z$ boson and the ALP,  $pp\to Za ~(Z\to l^+~l^-), (a\to\gamma\gamma)$, 
is depicted in the left panel of Fig.~\ref{fig:ppzax_diagram}. The production cross section $\sigma$ 
including the branching ratios versus the ALP mass $M_a$ is shown in 
Fig.~\ref{fig:ppzax_Xsection}. In Fig.~\ref{fig:scanCWCB}, we show the 2-D plot of the cross sections as a function of 
$(C_{WW}, C_{BB})$ for $f_a=1$~TeV and $M_a = 10$ GeV, $100$ GeV. We mark the cross point of $C_{WW}=2, C_{BB}=1$
in the figure, which are our benchmark values. The cross section simply scales as $1/f_a^2$ and would be of similar 
pattern for other values of $M_a$.
  
To delve into the physics behind this simulation, the photon propagator is specifically 
employed to scrutinize the coupling constant $g_{aZ \gamma}$, 
while the $Z$ propagator is utilized to investigate the coupling constant 
$g_{aZZ}$. This detailed analysis helps in understanding the behavior and 
interactions of the ALP with the electroweak gauge bosons in the given parameter space.
        
In the detection simulation, we opt for the \texttt{delphes\_card\_ATLAS.tcl} and employ the jet 
angle parameter $R = 0.4$ for clustering jets using FastJet~\cite{FastJet:2011} 
with the $anti-k_T$ algorithm~\cite{anti_kT:2008}. To enhance our analytical capabilities, 
we also calculate Nsubjettiness~\cite{Nsubjettiness:2010} in FastJet.

\subsubsection{Background analysis}\label{sec:Exp_Z_bg}

The final state is $l^+l^- \gamma\gamma$ and thus the predominant backgrounds are (i)
$pp \to l^+l^-\gamma\gamma$ (referred to as $ll\gamma\gamma BG$), (ii) $pp \to  l^+l^- \gamma j$ (referred to as $ll\gamma j BG$) with the jet faking a photon with a fake rate $f_{j \to \gamma} \simeq 5 \times 10^{-4}$ \cite{ATLAS:2017muo}, and (iii) $pp \to l^+l^-j$ (referred to as $lljBG$). Note that the $j$ here is the parton-level jet produced by \texttt{MadGraph5aMC@NLO} including gluon and light quarks, followed by showering. In the detector simulation level, the jets are clustered by FastJet with anti-$k_T$ algorithm. The first background is irreducible 
while in the second background the jet can fake a photon with a fake rate $f_{j\to\gamma} \sim 5\times 10^{-4}$ \cite{ATLAS:2017muo}. In the third background, the jet can fake a photon-jet for the case of light $M_a$. 
The importance of incorporating the $lljBG$ background lies in the potential for the photon pair
from a low-mass ALP to exhibit behavior akin to a jet when the angular separation is sufficiently small. A cutoff is imposed at the ALP mass $M_a=25~\rm GeV$, below which the first and the second backgrounds have to be taken into account in the signal-background analysis. Conversely, if the ALP mass is heavier than 25 GeV cutoff, the first and the third backgrounds is considered in the analysis.

Initial computations involve the evaluation of signal $s$ and background event $b$ rates 
at $\sqrt{s}=14~\rm{TeV}$, as described by the following equation:
    \begin{equation}\label{eq:5.1}
        s,b = \sigma_{s,b} \times \frac{N_{selected}}{N_{sim}} \times \mathcal{L}.
    \end{equation}
Here, $\sigma_s$ and $\sigma_b$ denote the cross-sections of signal and background, respectively,
including decay ratios. The ratio $\frac{N_{\text{selected}}}{N_{\text{sim}}}$ represents the selection efficiency, and $\mathcal{L}$ signifies the integrated luminosity. 
Figure \ref{fig:ppzax_Xsection} illustrates the cross 
sections of the signal process along with the corresponding backgrounds. 
The simulation involves the generation of $N_{\text{sim}} = 10^4$ signal events, $N_{\text{sim}} = 10^5$ for $ll\gamma \gamma BG$ and $ll\gamma jBG$ background events. For $lljBG$ background events, to maintain the accuracy, we generate $10^6$ events for this background channel. Additionally, two distinct integrated luminosities are considered, namely $300~\rm{fb^{-1}}$ (current run) and $3000~\rm{fb^{-1}}$ (high-luminosity run).

\subsubsection{Selection Procedures}\label{sec:Exp_Z_sel}

{\it $M_a > 25~\rm{GeV}$}

To minimize background event rates in the analysis, we examine the kinematic 
properties between the signal and backgrounds to establish a set of useful selection cuts.
Within the ALP mass region  $M_a > 25~\rm GeV$, we implement the 
following event selection cuts:
    \begin{itemize}
        \item two photon selection
        \item two lepton selection
        \item $80~\text{GeV}< M_{ll} < 100~\text{GeV}$
        \item $p_{T_{\gamma \gamma}}>80~\text{GeV}$
        \item $0.9M_a < M_{\gamma \gamma} < 1.1M_a$
    \end{itemize}

    \begin{figure}
        \centering
        \includegraphics[width=\textwidth]{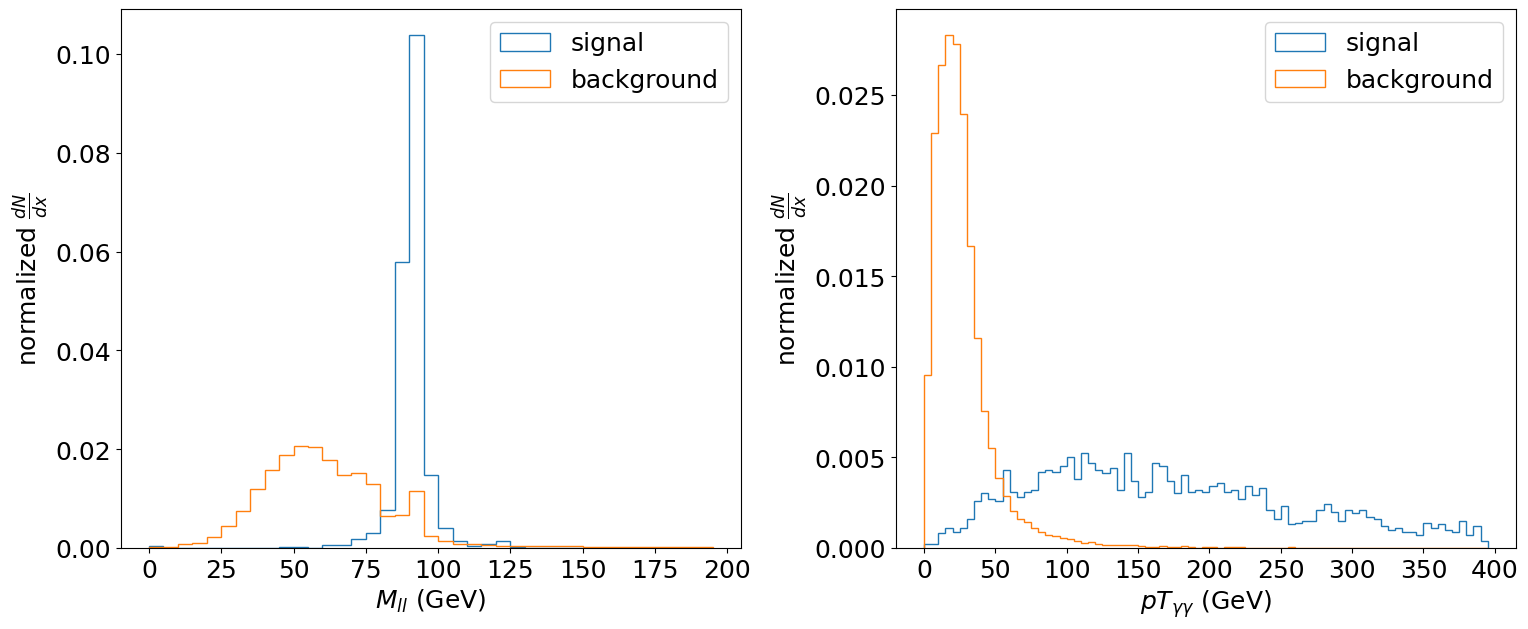}
        \caption{Normalized differential distribution $d N/ d M_{ll}$ (left panel) and 
        normalized differential
        distribution $d N / d p_{T_{\gamma\gamma}}$ (right panel) 
        for the signal $pp \to Za \to (l^+ l^-) (\gamma\gamma)$ 
        and $ll\gamma \gamma$BG with $M_a = 100$ GeV.
        }
        \label{fig:ppzax_kinematic}
    \end{figure}
    
 The final state consists of two isolated photons and two charged leptons; 
 therefore, it is imperative to apply the first two aforementioned cuts during the 
 selection procedures. Furthermore, the third selection cut aims to identify an 
 outgoing $Z$ boson with a mass of $M_Z=91.1876~\rm GeV$. 
 The distributions of the invariant mass $M_{ll}$  of charged lepton pair and
 the transverse momentum $p_{T_{\gamma\gamma}}$ of the photon pair
 are depicted in Fig.~\ref{fig:ppzax_kinematic}. It is apparent from the left panel of 
 Fig.~\ref{fig:ppzax_kinematic} that a significant portion of the background (BG) arises 
 from the $Z$ boson. Additionally, the right panel of Fig.~\ref{fig:ppzax_kinematic} 
 shows the significance of applying $p_{T_{\gamma\gamma}} > 80~\rm GeV$ cut 
 to mitigate the background events. The final cut is fixed by the ALP mass 
 window. For the upper and lower bounds of the mass, we opt for a selection of 
 10$\%$ on each side. The cut flow table for the ALP mass $M_a=100~\rm GeV$ is 
 presented in Table \ref{tab:z_selection_100}.

\begin{table}
 \centering
    \setlength{\tabcolsep}{15pt}
    \begin{tabular}{@{} p{0.3\textwidth} p{0.16\textwidth} p{0.16\textwidth} p{0.16\textwidth} @{}}
        \hline
        \textbf{Selection} & \textbf{Signal} & \textbf{$ll \gamma \gamma$BG} & \textbf{$ll \gamma j$BG} \\
        \hline
        Before cuts & 151948 & 29728 & 1048\\
        $N(\gamma)=2$ & 69243 & 12387 & 61.30\\
        $N(l)=2$ & 32152 & 5488 & 2.83\\
        $80~\rm GeV < M_{ll} < 100~\rm GeV$ & 29584 & 739 & 0.60\\
        $p{T_{\gamma \gamma}}>~\rm GeV$ & 24965 & 90 & 0.05\\
        $90~\rm GeV<M_{\gamma \gamma}<110~\rm GeV$ & 24707 & 13 & 0.02\\
        \hline
    \end{tabular}
    \caption{Cut flow for the signal ($pp\to Za$) with $Z\to l^+ l^-$ and 
    $a\to \gamma\gamma$ and the background process ($pp\to l^+l^- \gamma~\gamma$) and 
    ($pp \to l^+ l^- \gamma j$) 
    with $M_a=100~\rm GeV$, with $f_a=1~\text{TeV}$, $C_{WW}=2$, $C_{BB}=1$, and $C_g=g_{af}=0$.
    ``Before cuts" in the first row denotes the total number of events with only the parton-level cuts
    with the integrated luminosity of 300 fb$^{-1}$ calculated by Eq.~(\ref{eq:5.1}). 
    In $ll\gamma j$BG, we have applied the jet-fake rate  
    $f_{j\to\gamma}= 5 \times 10^{-4}$.}
    \label{tab:z_selection_100}
\end{table}

In Table.~\ref{tab:z_selection_100}, the first two cuts are essential for achieving 
the detection of the photon and the charged lepton pairs.
The selection cuts of $M_{ll}$ and $p_{T_{\gamma\gamma}}$ 
retain over 77\% of signal events and reduce background events to only 1.65\%. 
Finally, the ALP mass-window cut further reduces the background by a factor of 7.

\noindent
{\it $M_a < 25~\rm{GeV}$ }
    
In the low ALP mass region, where $M_a \leq 25~\rm GeV$, we implement the following cuts:
    \begin{itemize}
        \item At least one jet
        \item $min(\frac{E_{had}}{E_{EM}}) < 0.02$
        \item two lepton selection
        \item $80 ~{\rm GeV} < M_{ll} < 100~{\rm GeV}$
        \item $\frac{\tau_2}{\tau_1} < 0.05$
        \item $M_{\text{jet}}$ mass window
    \end{itemize}
As mentioned in Sec. \ref{sec:Exp_Z_bg}, the two photons decaying from a lower mass ALP form a photon-jet. 
Therefore, the first two selections aim at identifying the candidates for this type of ALP jets. 
Considering potential noise in hadron colliders, we accept multi-jet events. The candidate ALP jet 
should predominantly consists of photons, and thus the majority of its energy should be detected in 
the electromagnetic (EM) calorimeter. However, since there may be multiple jets in the event, a quantity, 
$\frac{E_{had}}{E_{EM}}$, is defined for each jet. We choose to select jets with the 
lowest $\frac{E_{had}}{E_{EM}}$ that is less than $0.02$.
The third and fourth selection cuts target at the $Z$ boson. 

The quantity $\tau_N$ represents the Nsubjettiness \cite{Nsubjettiness:2010, TASI_jet:2013}, which characterizes 
the substructure of a jet, defined by 
    \begin{equation}
    \tau_N = \frac{1}{d_0} \sum_{k} p_{T,k} \min \left\{ \Delta R_{1,k}, \Delta R_{2,k}, \ldots, \Delta R_{N,k} \right\}\\,
    \label{equ4.1}
    \end{equation}
where    
    \begin{equation*}
    \Delta R_{j,k} = \sqrt{(\Delta \eta)^2 + (\Delta \phi)^2},\qquad  {d_0} = \sum_{k} p_{T,k} R_0,
    \end{equation*}  
where $p_{T,k}$ is the transverse momentum of the $k^{\rm th}$ constituent particle, 
$\Delta \eta$ and $\Delta \phi$ are the pseudorapidity and azimuthal angle between $j^{\rm th}$ 
candidate subjet and $k^{\rm th}$ constituent particle, respectively, and $R_0$ is 
the characteristic jet radius, which we set to be $R_0=0.4$ in FastJet.
One identifies $N$ candidate subjets, after which $k$ iterates over all the other constituent 
particles in the jet to calculate this quantity. A jet with $\tau_N=0$ indicates that there 
are only $N$ or fewer constituent subjets in the jet, while a jet with $\tau_N>0$ indicates 
the presence of additional radiation outside the $N$ candidate subjets.
The ratio $\frac{\tau_N}{\tau_{N-1}}$ is a useful parameter to determine if a jet is composed of $N$ substructures. 
The ideal ALP jet, consisting of two photons without any other initial radiation, will result in $\tau_2$ being 
zero but with a non-zero $\tau_1$. The final selection involves the jet mass. We focus on the ALP jet with the 
lowest $\frac{E_{EM}}{E_{had}}$ as selected in the second cut. Instead of employing a symmetric mass window, 
the upper bound used is twice of the lower bound. The asymmetric mass window is motivated by the acceptance of
minimal initial radiation, which may contribute additional mass to the jet. The selection of $M_{\text{jet}}$
for various ALP mass window is listed in Table \ref{z_mass_selection_table}.

Taking $M_a=10~\rm GeV$ as an example, Table \ref{tab:z_selection_10} presents the cut flow.
One of the key cuts in Table~\ref{tab:z_selection_10} is the second one, which significantly 
reduces the contribution from hadronization jets in favor of ALP-induced jets. Additionally, 
the Nsubjettiness ratio cut effectively filters out jets resulting from the decay of light mesons. 
As a result, after applying these cuts, no background events pass the ALP mass selection.   

\begin{table}
    \centering
    \setlength{\tabcolsep}{15pt}
    \begin{tabular}{@{} p{0.35\textwidth} p{0.35\textwidth} @{}}
        \hline
        \textbf{$M_a\;(\rm GeV)$} & \textbf{$M_{\text{jet}}$ selection $(\rm GeV)$} \\
        \hline
        25 & $22.5 < M_{\text{jet}} < 30$\\
        20 & $18 < M_{\text{jet}} < 24$\\
        10 & $9 < M_{\text{jet}} < 12$\\
        5 & $4.5 < M_{\text{jet}} < 6$\\
        1 & $0.5 < M_{\text{jet}} < 2$\\
        \hline
    \end{tabular}
    \caption{$M_{\text{jet}}$ mass window for the low ALP mass region}
    \label{z_mass_selection_table}
\end{table}

    \begin{table}
    \centering
    \setlength{\tabcolsep}{15pt}
    \begin{tabular}{@{} p{0.32\textwidth} p{0.15\textwidth} p{0.15\textwidth} p{0.15\textwidth} @{}}
        \hline
        \textbf{Selection} & \textbf{Signal} & \textbf{$ll \gamma \gamma$ BG} & \textbf{$llj$ BG} \\
        \hline
        Before cuts & 426413 & 29728 & 164921970\\
        $N(jet) \geq 1$ & 356610 & 17327 & 139649327\\
        $min(\frac{E_{had}}{E_{EM}}) < 0.02$ & 267532 & 8150 & 26141121\\
        $N(l)=2$ & 88523 & 1169 & 649627\\
        $80~\rm GeV < M_{ll} < 100~\rm GeV$ & 81957 & 181 & 460297\\
        $\frac{\tau_2}{\tau_1} < 0.05$ & 62811 & 46 & 36613\\
        $9~\rm GeV < M_{jet} < 12~\rm GeV$ & 48995 & 0 & 0\\
        \hline
    \end{tabular}
    \caption{
    Cut flow for the signal process ($pp\to Za$) and background processes ($pp\to ll\gamma\gamma$) 
    and ($pp\to llj$) with $M_a=10~\rm GeV$, with
    $f_a=1~\rm{TeV}$, $C_{WW}=2$, $C_{BB}=1$, and $C_g=g_{af}=0$. 
    ``Before cuts" in the first row denotes the total number of events with only parton-level cuts
    with the integrated luminosity of 300 fb$^{-1}$ calculated by Eq.~(\ref{eq:5.1}). 
     }
    \label{tab:z_selection_10}
\end{table}


\subsection{$p p \to W a$ with $W \to l \nu_l$ and $a \to \gamma \gamma$}\label{sec:Exp_W}
Another useful production channel is $Wa$ production followed by the leptonic decay of the $W$
boson and $a\to \gamma\gamma$: $pp \rightarrow W^{\pm}a \left( W^{\pm} \rightarrow l^{\pm} \nu_{l} \right), 
\left( a \rightarrow \gamma\gamma \right)$, as illustrated in the right panel of Figure.~\ref{fig:ppzax_diagram}

\subsubsection{Signal Events }\label{sec:Exp_W_sig}

\begin{figure}[h!]
        \centering
        \includegraphics[width=0.75\textwidth]{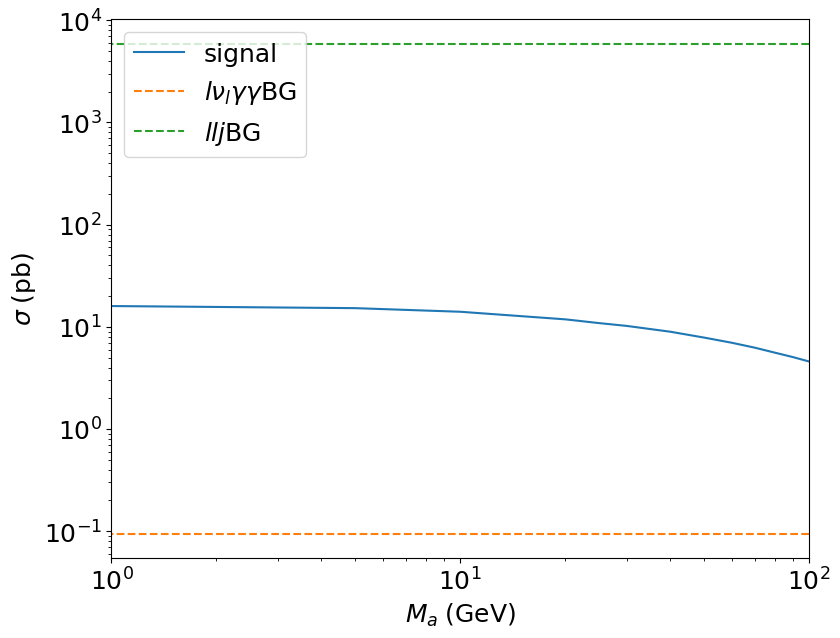}
        \caption{
        Production cross sections for $pp \to W^{\pm}a \left( W^{\pm} \to l^{\pm} \nu_{l} \right), \left( a \to \gamma\gamma \right)$ with
        $ l = e,\mu$
        versus $M_a$ at $\sqrt{s} = 14$ TeV, including 
        branching ratios. Here we also show $l\nu_l \gamma\gamma$BG and $l\nu_l j$BG. }
        \label{fig:ppwax_Xsection}
\end{figure}

Similar to the case of $pp\to Z a$, we set the parameters $f_a=1~\rm{TeV}$, $C_{WW}=2$, $C_{BB}=1$, 
and $C_g=g_{af}=0$, and the ALP decay width is set to "auto" in the \texttt{parm\_card.dat} of 
\texttt{MadGraph5aMC@NLO}, and its mass range is from $M_a =1 - 100 {\rm GeV}$. 
The production cross section of the signal including the branching ratios is showed in Fig.~\ref{fig:ppwax_Xsection}.

\subsubsection{Background analysis}\label{sec:Exp_W_bg}
The final state consists of a charged lepton and missing energy from $W$ decay and a pair of photons 
from the ALP decay. We consider two major backgrounds: (i) $pp \rightarrow l^{\pm} \nu_l \gamma\gamma$ 
(referred to as $l\nu _l\gamma\gamma BG$), (ii) $pp \to l^{\pm} \nu_l \gamma j$ (referred to as $l\nu _l\gamma j BG$) with the jet faking a photon with a fake rate $f_{j\to \gamma} \simeq 5 \times 10^{-4}$, and (iii) $pp \rightarrow l^{\pm} \nu_l j $ 
(referred to as $l\nu_l j BG$ ). 
Similar to the $Za$ process, we use a cutoff of 25 GeV, such that different background consideration 
is applied to the case $M_a < 25$~ GeV and the case $M_a > 25$~GeV.

\subsubsection{Selection procedures}\label{sec:Exp_W_sel}

{\it $M_a > 25~\rm{GeV}$}

\begin{figure}[h!]
        \centering
        \includegraphics[width=\textwidth]{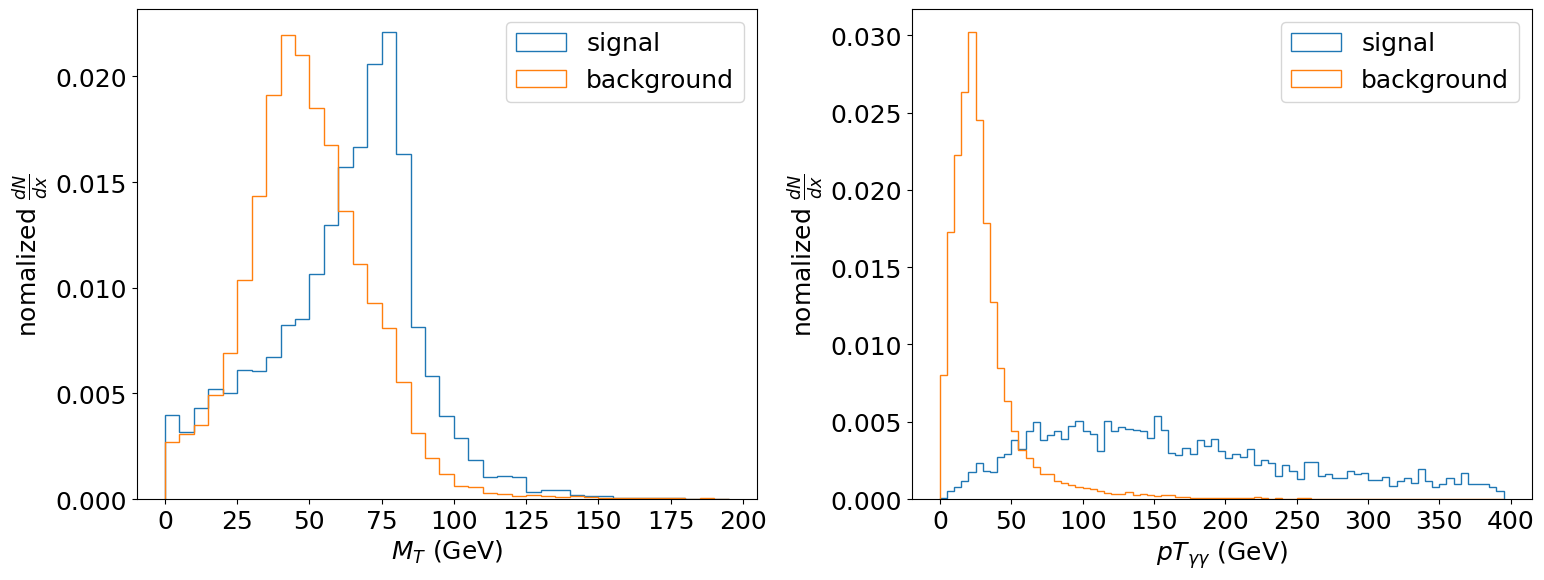}
        \caption{Normalized distributions of $dN/ d M_T$ and 
        $dN/d p_{T_{\gamma\gamma}}$ for $pp \to W^\pm a \to (l^\pm \nu_l) (\gamma\gamma)$ 
        for ${M_a}=100~\rm GeV$. The cuts associated 
        with $N(\gamma)=2$ and $N(l)=1$ have been applied.
        }
        \label{fig:ppwax_kinematic}
\end{figure}

To differentiate between the signal and background events, we apply the following cuts 
for the ALP mass range of $25~\text{GeV} < M_a \leq 100~\text{GeV}$.
    \begin{itemize}
        \item two photon selection
        \item one lepton selection
        \item $p_{T_{\gamma \gamma}}>50~\text{GeV}$
        \item $M_T>58~\text{GeV}$
        \item $0.9M_a < M_{\gamma \gamma} < 1.1M_a$
    \end{itemize}
The first two cuts in Table~\ref{tab:w_selection_100} identify one charged lepton and two photons in the final state.
Furthermore, as illustrated in Fig.~\ref{fig:ppwax_kinematic},  $p_{T_{\gamma \gamma}}$ is a 
useful kinematical variable to suppress the background. We choose
$p_{T_{\gamma \gamma}}>50~\text{GeV}$ . 
Since there is missing energy in the final state due to the leptonic decay of the $W$ boson, 
we utilize the transverse mass ${\displaystyle M_{T}=\sqrt{(E_{T,\l}+E_{T,mis})^{2}-({\vec {p}}_{T,\l}+
{\vec {p}}_{T,mis})^{2}}}$ to align with the Jacobian peak of the $W$ boson decay.
We choose $M_T > 58$~GeV to differentiate from the background (as seen in Fig \ref{fig:ppwax_kinematic}). 
The mass window of the final selection cut depends on the ALP mass. 
Given that the average peak of  $M_{\gamma \gamma}$ typically aligns with $M_a$ and the peak 
width diminishes as the ALP mass decreases, we opt for ${\pm}10\% $ of $M_a$ as the upper and lower limits 
for the mass window.
    
\begin{table}[h!]
    \centering
    \setlength{\tabcolsep}{15pt}
    \begin{tabular}{@{} p{0.3\textwidth} p{0.16\textwidth} p{0.16\textwidth} p{0.16\textwidth} @{}}
        \hline
        \textbf{Selection} & \textbf{Signal} & \textbf{$l\nu_l\gamma \gamma$ BG} & \textbf{$l\nu_l\gamma j$ BG}\\
        \hline
        Before cuts & 1375200 & 28311 & 5268\\
        $N(\gamma)=2$ & 617190 & 11435 & 152.2\\
        $N(l)=1$ & 402521 & 6954 & 12.3\\
        $p{T_{\gamma \gamma}}>50~\text{GeV}$ & 372542 & 906 & 3.5 \\
        $M_T>58~\text{GeV}$ & 232821 & 441 & 1.8\\
        $90 < M_{\gamma \gamma} < 110$ & 230208 & 48 & 0.16\\
        \hline
    \end{tabular}
    \caption{Cut flow for the signal $pp\to W^\pm a$ and the background
    ($pp\to l\nu_l\gamma\gamma$) and ($pp \to l\nu_l \gamma j$)  with $M_a=100~\rm GeV$, with couplings $f_a=1~\rm{TeV}$, 
    $C_{WW}=2$, $C_{BB}=1$, and $C_g=g_{af}=0$. 
  ``Before cuts” in the first row denotes the total number of events with only the parton-level cuts
    computed using 
    Eq.~(\ref{eq:5.1}), with the signal and background cross sections given in 
    Fig.~\ref{fig:ppwax_Xsection} and the luminosity set at $\mathcal{L}=300~\rm{fb^{-1}}$. In $l \nu_l \gamma j$BG, we have applied the jet-fake rate 
    $f_{j\to\gamma} = 5 \times 10^{-4}$.}
    \label{tab:w_selection_100}
\end{table}

\noindent
{\it $M_a < 25~\rm{GeV}$}

In the low ALP mass region, where $M_a\leq25~\rm GeV$, we implement the following cuts:
    \begin{itemize}
        \item At least one jet
        \item $min(\frac{E_{had}}{E_{EM}}) < 0.02$
        \item one lepton selection
        \item $M_T>58~\text{GeV}$
        \item $\frac{\tau_2}{\tau_1} < 0.05$
        \item $M_{\rm{jet}}$ mass window
    \end{itemize}
    
\begin{table}[h!]
    \centering
    \setlength{\tabcolsep}{15pt}
    \begin{tabular}{@{} p{0.32\textwidth} p{0.15\textwidth} p{0.15\textwidth} p{0.15\textwidth} @{}}
        \hline
        \textbf{Selection} & \textbf{Signal} & \textbf{$l\nu_l \gamma \gamma$ BG} & \textbf{$l\nu_l j$ BG} \\
        \hline
        Before cuts & 4218000 & 28311 & 1766100000\\
        $N(jet) \geq 1$ & 3289618 & 13415 & 1468625180\\
        $min(\frac{E_{had}}{E_{EM}}) < 0.02$ & 2782193 & 6824 & 250477133\\
        $N(l)=1$ & 1412608 & 2235 & 26117087\\
        $M_T>58~\text{GeV}$ & 888733 & 765 & 16493608\\      
        $\frac{\tau_2}{\tau_1} < 0.05$ & 595160 & 200 & 665820\\
        $9~\rm GeV < M_{jet} < 12~\rm GeV$ & 461449 & 2 & 1766\\
        \hline
    \end{tabular}
    \caption{ Cut flow for the signal $pp\to W^{\pm}~a$ and backgrounds 
    $pp\to l\nu_l \gamma\gamma$ and $pp\to l\nu_l j$ with $M_a=10~\rm GeV$, 
    featuring couplings $f_a=1~\rm{TeV}$, $C_{WW}=2$, $C_{BB}=1$, and $C_g=g_{af}=0$. 
    ``Before cuts” in the first row denotes the total number of events with only the parton-level cuts
    computed using Eq.~(\ref{eq:5.1}), with the signal and 
    background cross sections given in Fig.~\ref{fig:ppwax_Xsection} and the luminosity set at
    $\mathcal{L}=300~\rm{fb^{-1}}$.}
    \label{tab:w_selection_10}
\end{table}

 The decay of low-mass ALPs into two photons results in the formation of photon-jets. Thus, we have changed 
 the selection criteria from two photons to selecting at least one jet. To ensure that the selection
 includes jets composed of two photons, we select the jet with the smallest $\frac{E_{had}}{E_{EM}}$ 
 value, which must be under $0.02$. 
 The third and fourth selection cuts are similar to the case of $M_a > \rm 25~GeV$
 to identify the decay of the $W$ boson.
 The fifth selection cuts on the ratio of Nsubjettiness effectively minimizes the impact of the jet background. 
We employ an asymmetric mass window for the $M_{\text{jet}}$ cut, the same as the case of $Za$.
The cut flow table for $M_a=10~\rm GeV$ is presented in Table~\ref{tab:w_selection_10}.

\section{Numerical results}\label{sec:results}

    \begin{figure}[h!]
        \centering
        \includegraphics[width=\textwidth]{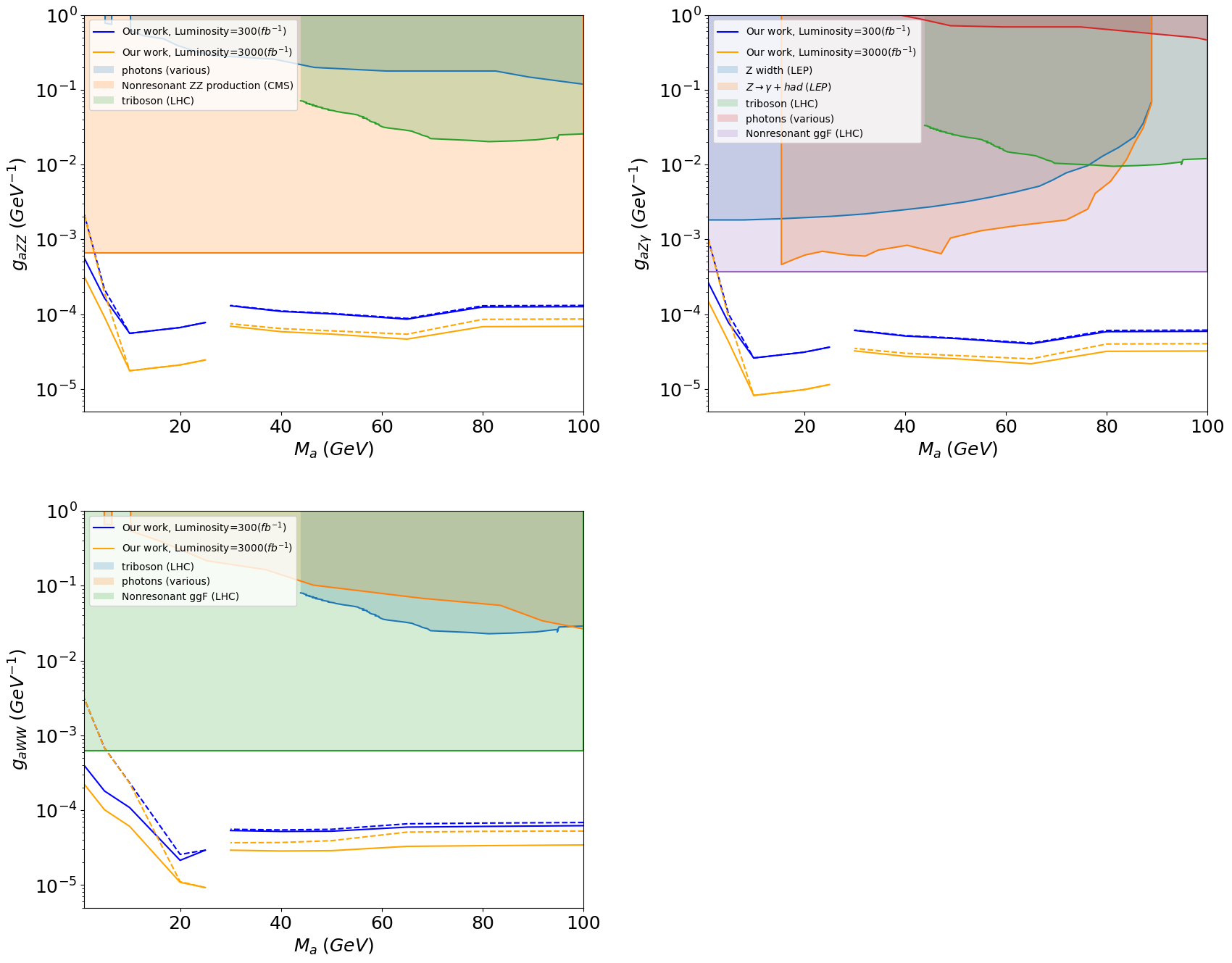}
     \caption{95\% C.L. exclusion regions for ALP-gauge couplings at the LHC ($\sqrt{s}=14$ TeV)
     with 300 fb$^{-1}$ (blue lines) and 3000 fb$^{-1}$ (orange lines) integrated luminosities. 
     We show the results with 10\% systematic uncertainty (dashed) and without 
     systematic uncertainty (solid).
    \textit{Top Left}: $g_{aZZ}$ resulting from $pp\to Za(Z\to l^+l^-)(a\to\gamma\gamma)$ . 
    \textit{Top Right}: $g_{aZ\gamma}$ from $pp\to Za(Z\to l^+l^-)(a\to\gamma\gamma)$. 
    \textit{Bottom Left}: $g_{aWW}$ resulting from $pp\to Wa(W\to l~\nu_l)(a\to\gamma\gamma)$.
 The other existing limits are described in Sec.~\ref{sec:Constrints}.
  }       \label{fig:final_result}
    \end{figure}

    \begin{figure}
        \centering
        \includegraphics[width=\textwidth]{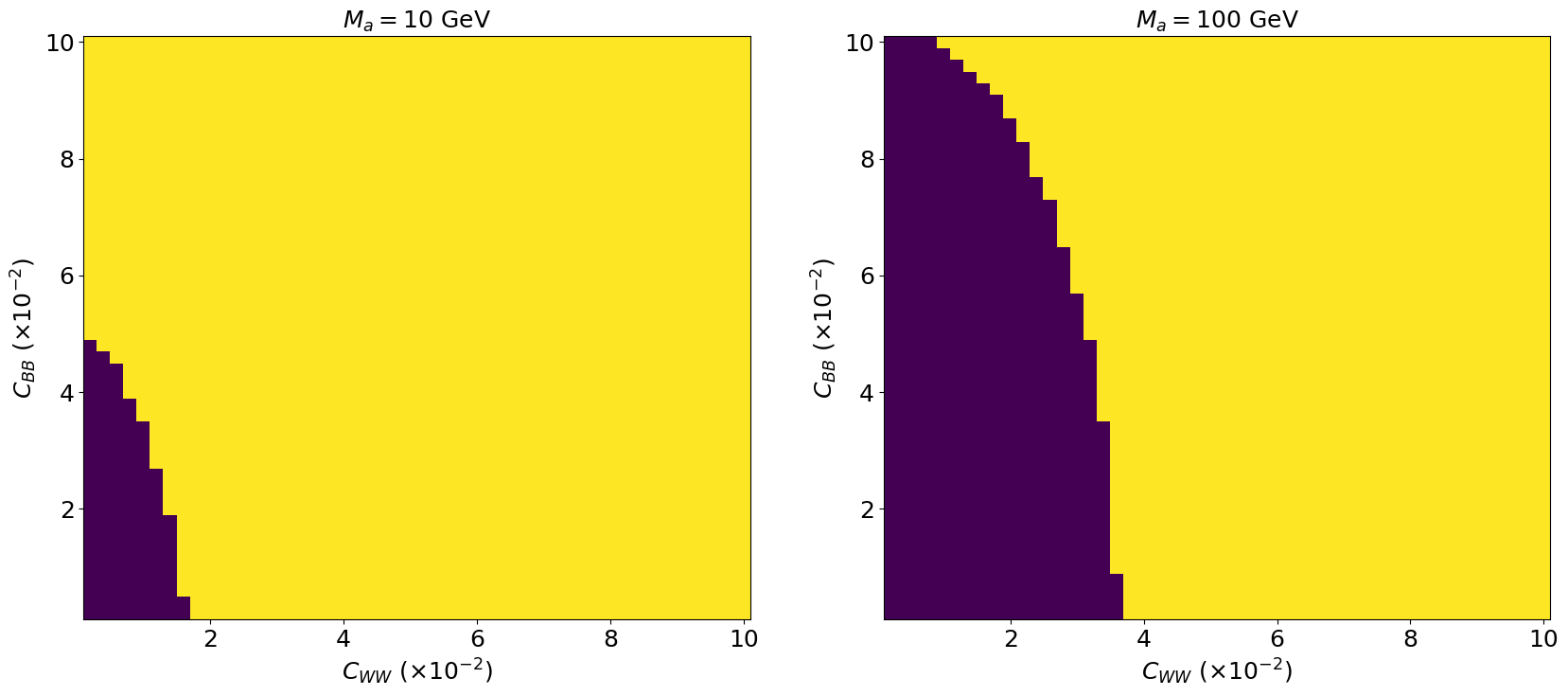}
        \caption{Sensitivity plot in the plane of $(C_{WW}, C_{BB} )$ for $M_a = 10$ GeV (left panel) and for $M_a = 100$ GeV (right panel).  We fix $f_a = 1$ TeV. The bright region on the left panel is for the number of signal events $>3$ in case with zero background, while that on the right panel is for significance $Z>2$. Note that the signal cross section scales as $1 / f_a^2$.}
        \label{fig:95CL}
    \end{figure}

In this section, we are going to derive the sensitivity reach on the ALP-gauge couplings
$g_{aZZ}$, $g_{aZ\gamma}$, and $g_{aWW}$ using the processes $pp \to Za \to (l^+ l^-) (\gamma\gamma)$
and $pp \to Wa \to (l \nu) (\gamma\gamma)$ at the 14 TeV LHC with integrated luminosities 
of 300 fb$^{-1}$ and 3000 fb$^{-1}$. 
In the last section, we have illustrated the signal events rates for a choice of 
$f_a = 1$~TeV, $C_{WW}=2$, and $C_{BB}=1$ ($C_g= C_{af}=0$). We use a simple scaling to 
estimate the sensitivity reach. 
The correlation between the number of events and the ALP-gauge couplings is 
expressed through Eqs.~(\ref{Eq.3}) -- (\ref{Eq.7}):
    \begin{equation}\label{Eq:5.2}
        s \propto \; f_a^{-2} \; \propto \; g_{aZZ}^2 ,\;\; g_{aZ\gamma}^2 , \;\; g_{aWW}^2 \;.
    \end{equation}
The 95\% confidence level (C.L.) sensitivity for the ALP-gauge couplings
can be determined by requiring the significance $Z > 2$, defined by \cite{Cheung:2023nzg,Arhrib:2019ywg}:
    \begin{equation}\label{Eq:5.3}
        Z = \sqrt{2\left[(s+b)\ln\left(\frac{(s+b)(b+\sigma_b^2)}{b^2+(s+b)\sigma_b^2}\right) - \frac{b^2}{\sigma_b^2}\ln\left(1+\frac{\sigma_b^2 s}{b(b+\sigma_b^2)}\right)\right]} \;,
    \end{equation}
where $s$ and $b$ represent the numbers of signal and background events, respectively. 
Additionally, $\sigma_b$ denotes the systematic uncertainty associated with the SM background estimation. 
We consider two scenarios for $\sigma_b= 0\%$ and 10\% of background events. 
For the process in which all background events are excluded, as illustrated in Table~\ref{tab:ppzax_app} and \ref{tab:ppwax_app} of appendix~\ref{sec:app}, the 95\% C.L. is estimated by requiring 3 signal events.

Figure \ref{fig:final_result} displays our final result for the limits on the ALP-gauge couplings 
alongside with several existing constraints. The upper-left and upper-right panels depict the 
limits on $g_{aZZ}$ and $g_{aZ \gamma}$ obtained from the $pp \to Za$ process, 
while the bottom panel represents the limit on $g_{aWW}$ derived from the $pp \to Wa$ process.
The shaded regions are those excluded by the current constraints. Our results are denoted by the
blue curve for the 300 fb$^{-1}$ integrated luminosity and by the orange curve for the 3000 fb$^{-1}$.

In the high-mass region, $25~\text{GeV} < M_a < 100~\text{GeV}$, our sensitivity curves demonstrate an 
improvement of approximately one order of magnitude compared to the current limits. 
Despite that the signal process has a larger cross section than the background processes, 
the enhancement is not only attributed to this factor, but also the selection criteria, which  
also play a pivotal role.
We have focused  on the diphoton decay mode of the ALP, and notably, there are no SM particles 
decaying into diphotons within the mass range of 25 GeV to 100 GeV. 

In the low-mass region, $1~\rm{GeV} < \mathit{M_a} < 25~\rm{GeV}$, even with 
an additional background process with substantial cross-sections, our results 
consistently exhibit an improvement of one to two orders of magnitude compared to the current 
limits, except for the case of $M_a=1~\rm{GeV}$. 
In handling the features of jet substructure, we have adopted a more stringent approach, thus 
resulting in the exclusion of jet backgrounds. The sudden degradation in the limit at 
$M_a=1~\rm{GeV}$ can be attributed to the diphoton decay mode of the neutral pion, 
where the reconstructed mass aligns with our selection range.

For completeness, we show in Fig.~\ref{fig:95CL} the sensitivity region of 95\% C.L. in the plane
of $(C_{WW}, C_{BB})$ with $f_a = 1$~TeV for $M_a = 10$ GeV (left panel) and $M_a=100$ GeV (right
panel). The choice of $M_a=10,100$ GeV corresponds to the two ranges of $M_a$ in our analysis with the corresponding cross-section shown in Fig. \ref{fig:scanCWCB} .

Since we have employed distinct signal selection criteria to address both heavier and lighter ALPs, 
a gap arises around the cutoff, $M_a=25~\rm{GeV}$. The best sensitivity is observed around 
$M_a=10~\rm{GeV}$, where $g_{aZZ} \approx 2 \times 10^{-5}~\rm{GeV^{-1}}$, $g_{aZ\gamma} \approx 8 \times 10^{-6}~\rm{GeV^{-1}}$, and $g_{aWW} \approx 8 \times 10^{-6}~\rm{GeV^{-1}}$ for a luminosity of $\mathcal{L}=3000~\rm{fb^{-1}}$. This is attributed to the scarcity of background events 
in the region $10~\rm{GeV} < \mathit{M_a} < 25~\rm{GeV}$ due to stringent selection criteria. 
Nevertheless, the signal cross sections, as depicted in Figs.~\ref{fig:ppzax_Xsection} and 
\ref{fig:ppwax_Xsection}, diminish with an increase in the ALP mass.

\section{Conclusions}\label{sec:conclusion}

In this study, we have investigated the sensitivity potential of the current run and the 
future High-Luminosity run of the LHC ($\mathcal{L}=300~\rm{fb^{-1}}~\text{and}~3000~fb^{-1}$) at 
the center-of-mass energy of $\sqrt{s}=14$ TeV, focusing on probing the dimensionful 
coupling constants $g_{aZZ}$, $g_{aZ\gamma}$, and $g_{aWW}$ associated with the axion-like particle.
This exploration was conducted through the processes $pp\to Za(Z\to l^+l^-)(a\to\gamma\gamma)$ and $pp\to Wa(W^\pm \to l^\pm ~\nu_l)(a\to\gamma\gamma)$. Our results demonstrated that these channels provide the most stringent bounds for the couplings $g_{aZZ}$, $g_{aZ\gamma}$, and $g_{aWW}$.

To maintain generality, we adopted the values $C_{WW}=2$ and $C_{BB}=1$, ensuring that the couplings $g_{aZZ}$, $g_{aZ\gamma}$, and $g_{aWW}$ are interrelated as shown in Eqs.~(\ref{Eq.4})-(\ref{Eq.7}) with all of them being non-zero. The analysis can be readily extended to explore independent coupling strengths.

In conclusion, our study has culminated in the presentation of a summary plot 
(Fig.~\ref{fig:final_result}) illustrating the sensitivity of $g_{aZZ}$, $g_{aZ\gamma}$, and $g_{aWW}$ 
achievable at the LHC, which are compared with existing constraints. Our estimations of the bounds on 
the axion-like particle gauge boson couplings versus the ALP mass from $M_a=1$ GeV to 100 GeV, and 
provide valuable insights for future experiments dedicated to the detection of ALPs.

\section*{Acknowledgment}
The work was supported in part by NSTC under the grant number
MOST-110-2112-M-007-017-MY3.

\begin{appendix}
\section{Event Rates}\label{sec:app}

In this appendix, we list the total number of signal and background events for various 
ALP masses from $M_a=1$ GeV to 100 GeV that \textit{before} and \textit{after} 
all the cuts mentioned in Sec.~\ref{sec:Exp}.
We show in Table~\ref{tab:ppzax_app} for $Za$ channel and in Table~\ref{tab:ppwax_app}
for $Wa$ channel.

\begin{table}[h!]
    \centering
    \setlength{\tabcolsep}{15pt}
    \begin{tabular}{ccccccc}
    \hline \hline
    $M_a~(\rm{GeV})$    &   \multicolumn{2}{c}{\bf Signal}  &\multicolumn{2}{c}{\bf $ll\gamma \gamma$BG+$ll\gamma j$BG}    &   \multicolumn{2}{c}{\bf $llj$BG}\\
    \cline{2-7}

        &   before  &   after   &   before  &   after   &   before  &  after\\
    \hline \hline
    100 &   151947  &   24707   &           &   12.80   &           &  \\
    80  &   184160  &   25211   &           &   12.78   &           &  \\
    65  &   214813  &   23522   &           &   7.13    &           &  \\
    50  &   253073  &   23207   &   30775   &   4.16    &   \multicolumn{2}{c}{-}\\
    40  &   284456  &   19286   &           &   3.87    &           &  \\
    30  &   321176  &   12205   &           &   2.97    &           &  \\
    \hline
    25  &   342005  &   25206   &           &   0       &           &  0\\
    20  &   366568  &   34311   &           &   0       &           &  0\\
    10  &   426413  &   48995   &   29728   &   0       & 164950770 &  0\\
    5   &   457138  &   48914   &           &   0.59    &           &  165\\
    1   &   472141  &   46175   &           &   29.43   &           &  18474\\
    \hline \hline
    \end{tabular}
    \caption{Total number of signal events for $pp \to Za$, followed by  
    $Z\to l^+ l^-, a \to \gamma \gamma$ and 
    background events of $ll\gamma\gamma$BG and $llj$BG for the mass range $M_a=1-100~\rm{GeV}$.    
    The number of events are calculated by Eq.~(\ref{eq:5.1}), where the cross sections of 
    signal and backgrounds are shown in Fig.~\ref{fig:ppzax_Xsection} and the integrated 
    luminosity is set at $\mathcal{L}=300~\rm{fb^{-1}}$.}
    \label{tab:ppzax_app}
\end{table}

\begin{table}[h!]
    \centering
    \setlength{\tabcolsep}{12pt}
    \begin{tabular}{ccccccc}
    \hline \hline
    $M_a~(\rm{GeV})$    &   \multicolumn{2}{c}{\bf Signal}  &\multicolumn{2}{c}{$l\nu_l \gamma \gamma$\bf BG +$l\nu_l\gamma j$\bf BG}    &   \multicolumn{2}{c}{$l\nu_l j$\bf BG}\\
    \cline{2-7}

       &   before  &   after   &   before  &   after   &   before  &  after\\
\hline \hline
100 &   1375200 & 230208 &           & 48.01  &           &      \\
80  &   1674900 & 244870 &           & 52.87  &           &  
\\
65  &   1992000 & 261151 &           & 57.09  &           &      \\
50  &   2357400 & 245170 &   33579   & 36.57  &           &      \\
40  &   2690700 & 220906 &           & 26.43  &           &      \\
30  &   3066000 & 193771 &           & 21.05  &           &      \\
\hline
25  &   3267000 & 224443 &           &  0      &            &  0  \\
20  &   3555000 & 290799 &           &  0.57   &            &  0  \\
10  &   4218000 & 461449 &   28311   &  1.98   & 1766100000 &  1766  \\
5   &   4575000 & 522007 &           &  6.79   &            &  17661  \\
1   &   4797000 & 469147 &           &  110    &            &  326729  \\
\hline \hline
    \end{tabular}
    \caption{Total number of signal events for 
    $pp \to W^\pm a$, followed by  $W^\pm\to l^\pm \nu_l, a \to \gamma \gamma$ and 
    background events of $l\nu_l \gamma\gamma$BG and $l\nu_l j$BG for the mass range $
    M_a=1-100~\rm{GeV}$.    
    The number of events are calculated by Eq.~(\ref{eq:5.1}), where the cross sections of 
    signal and backgrounds are shown in Fig.~\ref{fig:ppwax_Xsection} and the integrated 
    luminosity is set at $\mathcal{L}=300~\rm{fb^{-1}}$.}
    \label{tab:ppwax_app}
\end{table}

\end{appendix}

\end{document}